\documentclass[10pt,apl,twocolumn]{revtex4}

\usepackage[dvips]{graphicx}
\usepackage{amsmath}

\begin{document}

\title{Observation of photon noise by cold-electron bolometers}
\author{A.V. Gordeeva$^{1,2,3}$}
\email{anna.gord@list.ru}
\author{V.O. Zbrozhek$^1$}
\author{A.L. Pankratov$^{1,2,3}$}
\author{L.S. Revin$^{1,2,3}$}
\author{V.A. Shamporov$^{1,2,3}$}
\author{A.A. Gunbina$^1$}
\author{L.S. Kuzmin$^{1,4}$}

\affiliation{$^1$Nizhny Novgorod State Technical University n.a. R.E. Alekseev, GSP-41, Nizhny Novgorod, 603950, Russia \\
$^{2}$Institute for Physics of Microstructures of RAS,
GSP-105, Nizhny Novgorod, 603950, Russia\\
$^{3}$Lobachevsky State University of Nizhny Novgorod,
GSP-20, Nizhny Novgorod, 603950, Russia\\
$^{4}$Chalmers University of Technology, 41296, Gothenburg, Sweden}

\begin{abstract}
We have measured a response to a black body radiation and noise of the cold-electron bolometers. The experimental results have been fitted by theoretical model with two heat-balance equations. The measured noise has been decomposed into several terms with the help of theory. It is demonstrated that the photon noise exceeds any other noise components, that allows us to conclude that the bolometers see the photon noise. Moreover, a peculiar shape of the noise dependence on the absorbed power originates completely from the photonic component according to the theory. In the additional experiment on heating of the cryostat plate together with the sample holder we have observed nearly independence of the noise on the electron temperature of the absorber, which has provided another proof of the presence of the photon noise in the first experiment.
\end{abstract}

\maketitle

The current cosmological experiments apply very strict requirements to the detectors installed on telescopes.
The Cosmic Microwave Background (CMB) is still the target of many Cosmology experiments. The main goal of the current cosmology is the understanding if an inflationary process occurred when the universe was about $10^{-37}$s old. Many theoretical models predict that this process should have left footprints of it in the form of a very small polarized signal  called B-modes. Detecting this signal would provide important information about the primordial universe and the high energy physics.

The recent joined analysis of BICEP2 experiment and Planck satellite results confirm that BICEP2 didn't detect any B-modes \cite{PlanckBICEP2015}. This work has demonstrated the stringent necessity of developing sensitive multifrequency CMB experiments to correct the observations from the dust contributions. Cold-electron bolometers \cite{Kuzmin2002, Kuzmin2012} are promising detectors for cosmological applications, as they have all the qualities necessary to perform tasks: such as the high sensitivity to terahertz radiation and immunity to cosmic rays \cite{rad}.

The purpose of this work is experimental demonstration that cold-electron bolometers have an ultimate sensitivity, i.e. sensitive to the photon noise. Photon noise is fluctuations, inevitably present in any radiation due to the discrete nature of the photons. The photon noise turns into a voltage noise of the detector, multiplied by the volt-watt response $S_V$ of the receiver. Ideally, all the other components of the detector noise, including an internal noise, must be smaller than the photonic component. If this condition is met, the detector is limited by the photon noise. Experimental demonstration of the photon noise is necessary for the installation of this type of bolometers on telescopes.

In this paper we present the results of optical experiments with a parallel-series arrays of cold-electron bolometers (CEBs) \cite{Kuzmin2008, Tarasov2011}. In order to obtain the main characteristic of detectors -- the noise equivalent power (NEP), one needs to know the absorbed power. At least three different approaches are used to determine the absorbed power in cold electron bolometers:
\begin{enumerate}
  \item from electromagnetic properties of the device, if the efficiency of antenna absorption is known. This approach was used, for example, in \cite{Tarasov2011};
  \item from simplified heat-balance equation at zero bias, where only two terms are nonzero. It was used, for example, in \cite{Brient2014},\cite{Brient2016};
  \item from the full set of heat-balance equations and fitting of the whole IV-curves \cite{Mukhin_RU}.
\end{enumerate}
We have shown the limitations of the first method in \cite{Mukhin_RU}. The second method can give a misleading result due to ambiguity of the zero-bias point. For example in \cite{Kuzmin2004} a zero bias peak was observed. In this paper we use the power, found from the heat balance equations, to calculate responsivity $S_V$ and NEP. In addition we estimate the incoming power, radiated by the black body in one mode. The latter quantity limits the possible absorbed power from the top.

\begin{figure}[h]
\resizebox{\columnwidth}{!}{
\includegraphics{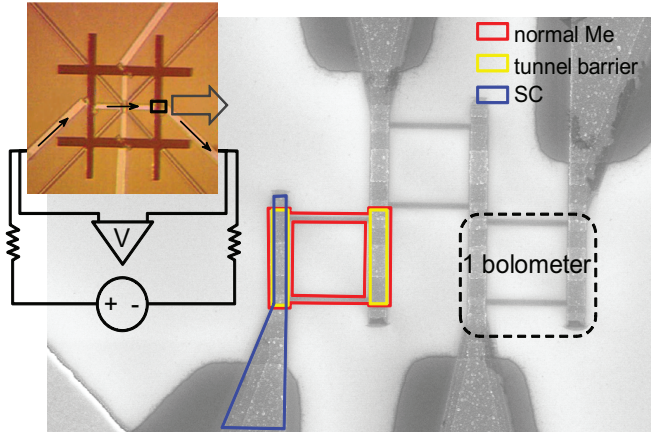}}
{\caption{{Main: Three bolometers in one slot of the antenna (SEM image).
Inset: cross-slot antenna in optical microscope. The array of bolometers is biased through bias resistors.}}
\label{fig01}}
\end{figure}

The sample is shown in Fig. \ref{fig01}. Four arrays, consisting of three bolometers each, are integrated into the cross slot antenna. The bolometers are connected in series for dc current and in parallel for high-frequency RF-current. The bolometer responses of opposite slots are summed up for dc read out. The details of the biasing circuitry is shown in the inset in Fig. \ref{fig01}. The presented design of cross-slot antenna is based on the original work for 550 GHz \cite{Zmuidzinas1998}. For our purpose the antenna has been scaled down to 350 GHz. The antenna is made of gold. The cold electron bolometer consists of normal metal absorber made of aluminium with suppressed superconductivity, aluminium oxide tunnel barrier and aluminium superconducting electrodes (SINIS junctions). All three layers are shown by colour frames in Fig. \ref{fig01}.

\begin{figure}[h]
\resizebox{\columnwidth}{!}{
\includegraphics{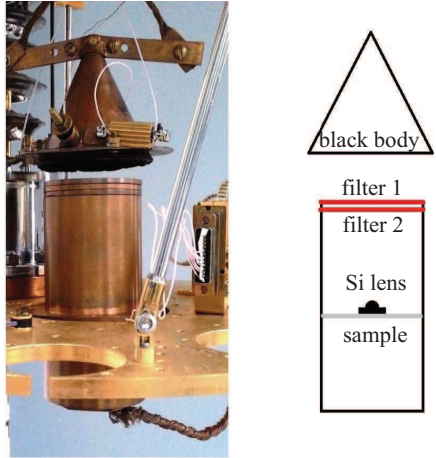}}
{\caption{{The view of experimental setup and its schematic representation.}}
\label{fig02}}
\end{figure}
The main parts of the experimental setup (Fig. \ref{fig02}) are: the antenna with bolometers coupled to a silicon lens with diameter 4 mm, the optical can with filters from one side and the source of black body radiation, attached to another cryostat plate. The sample in the can has been cooled down to 200 mK and illuminated by a black body (BB) with temperature varied from 2.7 K to 47 K.

Black body radiation was filtered by a set of filters fabricated in Cardiff University with bandwidth of 33 GHz and central frequency of 350 GHz. Their transmission is $10^{-8}$ from 500 to 850 GHz and $10^{-6}$ from 850 to 1200 GHz. The transmission above 1200 GHz was not measured. The equivalent filters were mounted on the optical window of the balloon cryostat of BOOMERanG \cite{Masi2006}.

\begin{figure}[h]
\resizebox{\columnwidth}{!}{
\includegraphics{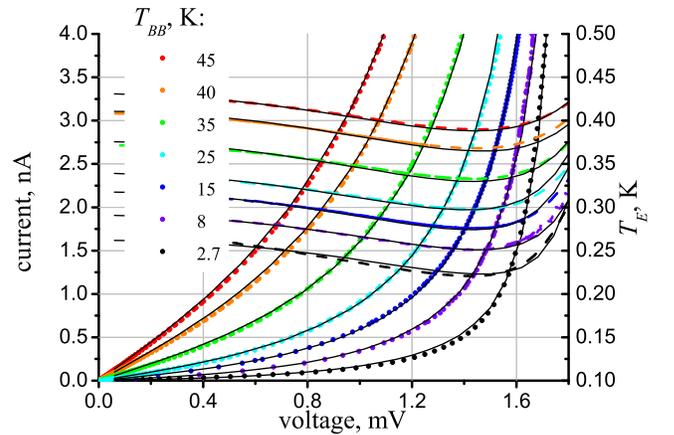}}
{\caption{Measured current (dots), the electron temperature, calculated from Eq. (1) (dashed curves) and the result of fitting from Eq. (3) (black solid curves) as functions of bolometer voltage for different black body temperatures $T_{BB}$.}
\label{fig03}}
\end{figure}

In Fig. \ref{fig03} the current-voltage characteristics of CEBs are shown for different BB temperatures. The figure also includes the electron temperature $T_E$ for each IV-curve, extracted from the tunneling current across NIS junction:
\begin{equation}
I = \frac{1}{e R_N}\int_{-\infty}^{\infty} \upsilon(\varepsilon)\left[ \frac{1}{exp(\frac{\varepsilon-eV}{\tau_E})+1} - \frac{1}{exp(\frac{\varepsilon}{\tau_S})+1}\right] d\varepsilon.
\label{Current}
\end{equation}
Here $\upsilon(\varepsilon)$ is a density of states in superconductor with Dynes parameter $\gamma$:
\begin{equation}
\upsilon(\varepsilon) = \left| \textrm{Re} \left[\frac{\varepsilon/\Delta - i \gamma}{\sqrt{(\varepsilon / \Delta - i \gamma)^2 - 1}}\right] \right|.
\end{equation}

Four parameters are required to find $\tau_E$ (normalized $T_E$) from Eq. (\ref{Current}): the critical temperature $T_c$ of superconductor, the normal state resistance of the junction $R_N$, the normalized temperature of electrons in superconductor $\tau_S$ and the parameter $\gamma$. The Eq. (\ref{Current}) has a vague dependence on $\tau_S$ and $\gamma$, so that only two parameters, $T_c$ and $R_N$, have to be measured in order to get accurate values of $T_E$ from IV-curves. For this sample we have used the following values: $T_c = 1.47$ K, $R_N = 8$ k$\Omega$ (for the array) and $\gamma = 5\cdot 10^{-5}$.

Two heat balance equations, for normal absorber \cite{Golubev} and for superconductor electrodes, respectively, can be written in the following form \cite{Mahashabde,O'Neil,Saleh}:
\begin{equation}
\begin{cases}P_N + P_{LOAD} + 2\beta P_S = P_{E-PH} + 2P_{COOL}\\
\epsilon(1-\beta)P_S = P_{E-PH,S}.\end{cases}
\end{equation}
Here  $P_N$ is the Joule heating of normal absorber,
$P_{COOL}$ is the cooling power of NIS junction,
$P_{LOAD}$ is the absorbed power of radiation (the source of photon noise),
$P_S$ is a power brought to superconductor by hot electrons from the absorber. Its fraction $2\beta$ returns to the normal metal, another fraction $\epsilon(1-\beta)$ heats electron subsystem in superconductor near the barrier, and the rest is transferred away and dissipates in the substrate.
$P_{E-PH} = \Sigma V (T^5_E - T^5_{PH}$) is the electron phonon coupling in normal metal.
$P_{E-PH,S} = 0.98  \Sigma_S V_S (T^5_S - T^5_{PH})e^{-\Delta(T_{S})/\tau_S}$ is the electron-phonon coupling in superconductor \cite{Timofeev2009}.

Some parameters are known from direct measurements or technological process: the absorber resistance $R_{abs} = 110$ $\Omega$, the absorber volume $V = 0.02 {\rm \mu m}^{-3}$ and the volume of superconducting electrode $V_S = 2.5 {\rm \mu m}^{-3}$. Whereas the material constants $\Sigma_N = 1.25\,\, {\rm nW} {\rm \mu m}^{-3}{\rm K}^{-5}$, $\Sigma_S = 0.3\,\, {\rm nW} {\rm \mu m}^{-3}{\rm K}^{-5}$, the backtunneling coefficient $\beta = 0.3$, degree of heating of superconducting electrons $\epsilon$ and the absorbed power $P_{LOAD}$ are found from the fit. The material constants are in agreement with literature data \cite{Giazotto2006} both for superconducting and normal aluminium.

The easiest for fitting and the most trustworthy from the point of view of the absorbed power is a region approximately from $0.1$ to $0.6$ of the voltage gap. There the absorbed power exceeds the other heating powers and therefore can be found with high accuracy. Closer to the gap the fitting is more tricky since the other heating terms start dominating. Nevertheless the results of our fitting with two heat balance equations are in good agreement with experiment for the whole range of voltages shown in Fig. \ref{fig03}. For example, the electron temperature, obtained from the fit, coincides with values, found directly from IV-curves using expression (\ref{Current}), with the accuracy better than 3 mK.

\begin{figure}[h]
\resizebox{\columnwidth}{!}{\includegraphics{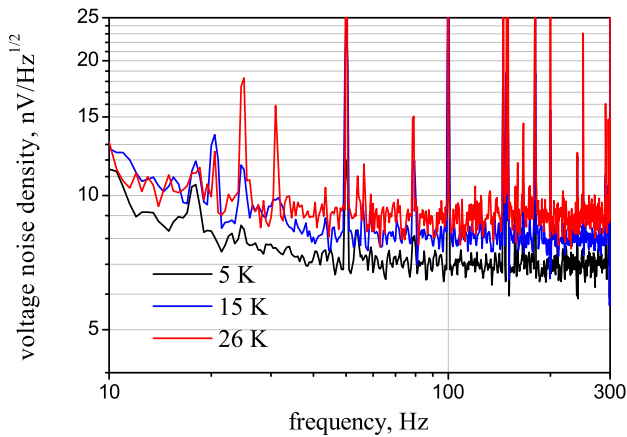}}
{\caption{Voltage noise at bias current 2 nA for various black body temperatures, $T_{PH} = 200$ mK.}
\label{SDf}}
\end{figure}

In Fig. \ref{SDf} the spectra of experimental voltage noise of CEBs biased at 2 nA are presented. One can see that change of the BB temperature increases the average noise level. The peaks at frequencies multiple to 50 Hz do not interfere the observation of the spectral flat region above 60-70 Hz.

\begin{figure}[h]
\resizebox{\columnwidth}{!}{\includegraphics{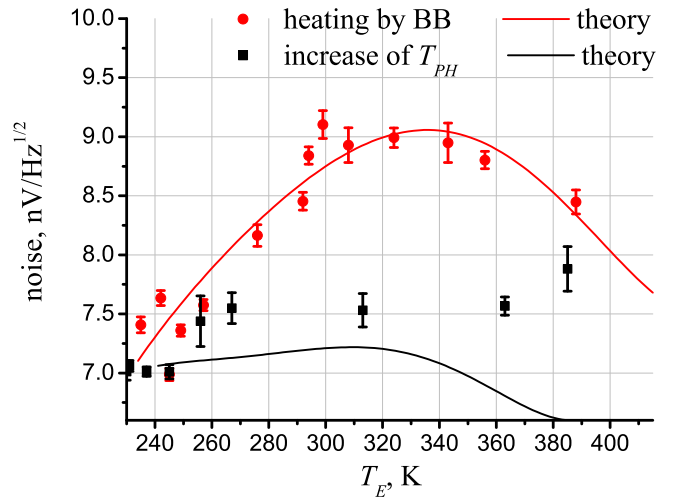}}
{\caption{Noise versus electron temperature for black body heating (red circles) at $T_{PH} = 200$ mK and plate heating (black squares). Solid curves - fitting results. Bias current is 2 nA. }
\label{fig05}}
\end{figure}

In Fig. \ref{fig05} the experimental noise at frequency 120 Hz versus electron temperature is presented at fixed bias current 2 nA. The electron temperature of the absorber has been increased in two ways: 1 - by heating of the sample holder and 2 - by heating of the black body. The former heats the phonons and then the electrons via week link between the two systems; the latter should heat the electrons much more efficiently since the power is absorbed directly by the electrons. The different character of noise behaviour in these two cases is a signature of the photon noise presence in the latter experiment. The different character of noise behavior in these two cases is a signature of the photon noise presence in the latter experiment. When $T_E$ in two experiments coincides, the differential resistance and the amplifier noise are the same as well, but the total noise is different.
In the first case the measured noise is higher than the theory due to temperature drifts, because the higher temperature set point of the coldest plate with the sample the harder to stabilize the cryostat.

\begin{figure}[h]
\resizebox{\columnwidth}{!}{\includegraphics{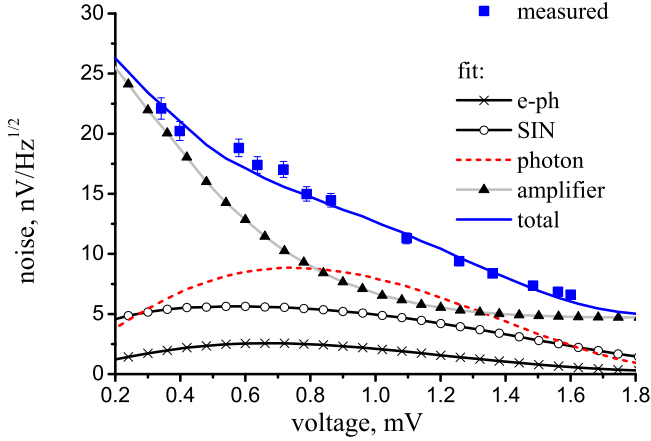}}
{\caption{Measured voltage noise at 120 Hz and different components of theoretical noise versus bolometer voltage $V$ for BB temperature 20 K, $T_{PH} = 200$ mK.}
\label{SDV}}
\end{figure}

The noise dependence on the bolometer voltage is presented in Fig. \ref{SDV}. The noise was measured at frequency 120 Hz and $T_{BB}=20$ K, which corresponds to 0.72 pW of absorbed power. At this power load the photon noise is significant and higher that the amplifier noise in the range (0.85 - 1.32)mV. In order to maximize the photon noise component in front of other components the operation point has to be slightly higher than the half gap.

The noise in cold-electron bolometers has several components: noise of electron-phonon interaction, SIN junction noise (includes shot noise of current through a tunnel junction, noise of heat flow and correlation between them), amplifier noise and photon noise. All components are written below in the listed order in terms of NEP:
\begin{equation}\label{NEP}
\begin{aligned}
& NEP_{amp}^2 = \frac{\delta V ^2 + (R \delta I)^2}{S_{V}^2},\\
& NEP_{E-PH}^2 = 10 k_B \Sigma V (T^6_E + T^6_{PH}),\\
& NEP_{SIN}^2 = \partial P^2 + \frac{\partial I^2}{(\partial I/\partial V \cdot S_V)^2} - 2\frac{\partial P \partial I}{\partial I/\partial V \cdot S_V},\\
& NEP_{photon}^2 = 2h\nu P_{LOAD} + \frac{P_{LOAD}^2}{\Delta \nu}.
\end{aligned}
\end{equation}
Here $\delta V = 4.7 \textrm{nV} / \sqrt{\textrm{Hz}}$ is the voltage noise and $\delta I = 12 \textrm{fA}/ \sqrt{\textrm{Hz}}$ is the current noise at 120 Hz of the amplifiers AD745. The amplifier noise was calibrated using a resistor and agrees with the passport data at 120 Hz: $5 \textrm{nV} / \sqrt{\textrm{Hz}}$ and $9 \textrm{fA}/ \sqrt{\textrm{Hz}}$.

\begin{figure}[h]
\resizebox{\columnwidth}{!}{\includegraphics{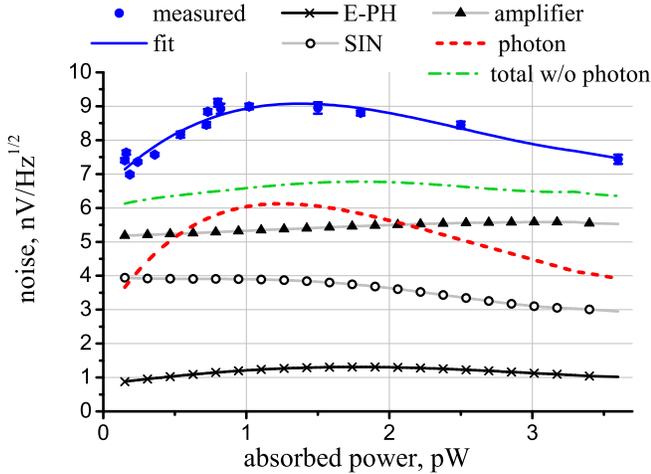}}
{\caption{Different components of noise versus absorbed power. Bias current 2 nA, $T_{PH} = 200$ mK.}
\label{fig07}}
\end{figure}
In Fig. \ref{fig07} the main result of the paper is shown: the experimental noise versus absorbed power for a fixed bias point $2$ nA decomposed in terms (\ref{NEP}) with the help of theoretical analysis. Theory and experiment are in good agreement. The photon noise component exceeds all other components separately. The least noise component, which can even be neglected, is the electron-phonon noise. The sum contribution of two other terms (SIN and amplifier) is slightly larger than the photon noise. Nevertheless, one can clearly see that the hilly shape of the total noise originates from the photonic component.

Let us consider the component of the photon noise $\delta V_{photon} = NEP_{photon} \cdot S_{V}$. Photon NEP has root dependence on power $P_{LOAD}$ for $\nu \gg \Delta \nu$, see Eq. (\ref{NEP}). This leads to a sharp increase in the photon noise from minimum absorbed power to $P_{LOAD} = 1 pW$, Fig. \ref{fig07}. Then the dependence $S_{V}(P_{LOAD})$ starts to play a decisive role. The experiment shows that starting from $P_{LOAD} = 1 pW$ the responsivity decreases by a factor of three resulting in a decline of the photon noise, Fig. \ref{fig07}. Thus, the experimental dependence of the noise in Fig. \ref{fig07}, confirmed by the theory, indicates that the bolometers see the photon noise at the background of system self-noise.
\begin{figure}[h]
\resizebox{\columnwidth}{!}{\includegraphics{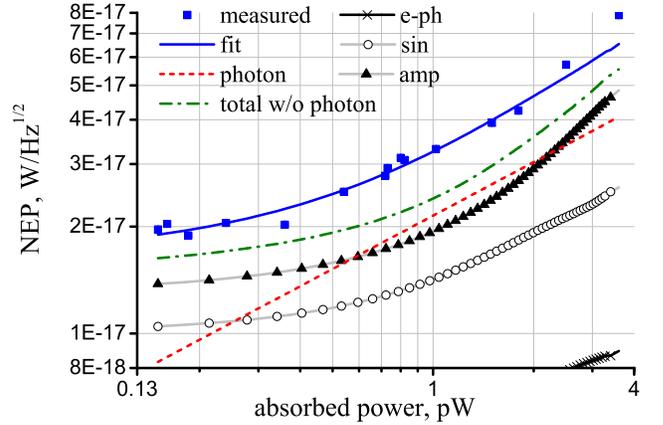}}
{\caption{Different components of NEP as a function of absorbed power. Bias current 2 nA, $T_{PH} = 200$ mK.}
\label{fig08}}
\end{figure}

The noise equivalent power versus power load for the considered bolometer array is shown in Fig. \ref{fig08}. As can be seen, this array performs closely to photon limited regime at power loads from 0.5 to 2 pW. Further decrease of the total noise can be made by decrease of the amplifier noise component. It can be made by change of read-out electronics to ac-scheme or to the correlation scheme as in \cite{Brient2014},\cite{Brient2016}.

We have found that the bolometers absorb less power (approximately by $30\%$) than has been expected from calculations of antenna efficiency \cite{Mukhin_RU}. Thus we do not encounter the problem of excess power discussed in \cite{Bueno2014}. But we have found that in order to fit IV-curves the
parameters $\beta$ and  $\epsilon$, responsible for returning power and for superconducting electrods
heating, has to be changed gradually with increase of $T_{bb}$. We assume this is due to nonequilibrium effects in superconducting leads, described, for example, in \cite{Vasenko2009}.

We have shown both experimentally and theoretically that the cold electron bolometers are sensitive to the photon noise. The responsivity of bolometers $S_V = dV/dP_{LOAD}$ at the bias point decreases from $5\cdot 10^8$ to $1.5\cdot 10^8$ V/W with the absorbed power. NEP increases from $2\cdot 10^{-17}\textrm{W Hz}^{-1/2}$ to $6\cdot 10^{-17}\textrm{W Hz}^{-1/2}$ correspondingly. The least ratio between NEP and NEP$_{PH}$ observed in our experiments is 1.5 for the absorbed power of 1-2 pW.

Authors wish to thank M.Tarasov for stimulating discussions and comments and Ernst Otto and Sumedh Mahashabde for help in the sample fabrication and Evgeniy Skorokhodov for making SEM image of the sample.

The work is supported by Russian Science Foundation (Project 16-19-10468).
The facilities of the Common Research Center "Physics and technology of micro- and nanostructures" of IPM RAS and of the Center of Cryogenic Nanoelectronics of NNSTU were used.

\end{document}